\documentclass[12pt]{iopart}
\usepackage{graphicx}
\usepackage{amsfonts}
\usepackage{amssymb}
\usepackage{multirow}
\newcommand{\dn}{\downarrow}
\newcommand{\up}{\uparrow}

\begin{document}

\title{Analysis of quantum spin models on hyperbolic lattices and Bethe lattice}

\author{Michal Dani\v{s}ka and Andrej~Gendiar}

\address{Institute of Physics, Slovak Academy of Sciences, SK-845~11, Bratislava, Slovakia}
\ead{michal.daniska@savba.sk and andrej.gendiar@savba.sk}

\date{\today}

\begin{abstract}
The quantum XY, Heisenberg, and transverse field Ising models on hyperbolic lattices are studied by means of the Tensor Product Variational Formulation algorithm. The lattices are constructed by tessellation of congruent polygons with coordination number equal to four. The calculated ground-state energies of the XY and Heisenberg models and the phase transition magnetic field of the Ising model on the series of lattices are used to estimate the corresponding quantities of the respective models on the Bethe lattice. The hyperbolic lattice geometry induces mean-field-like behavior of the models. The ambition to obtain results on the non-Euclidean lattice geometries has been motivated by theoretical studies of the anti-de Sitter/conformal field theory correspondence.
\end{abstract}


\maketitle

\section{Introduction}
Many analytical and computational techniques have been developed to study quantum spin models on two-dimensional (Euclidean) lattices. Among such techniques, let us mention the corner transfer matrix approach~\cite{Baxter}, the coordinate Bethe Ansatz~\cite{Bethe}, the algebraic Bethe Ansatz~\cite{Fad}, the vertex operator approach~\cite{AA}, including numerical algorithms based on tensor product states and tensor networks~\cite{Orus,TPS1,TPS2,TPS3,HOTRG}, all of which have been successfully applied to the description of the energy spectrum and matrix elements of local operators in either integrable lattice models and quantum spin chains or non-integrable quantum spin systems. However, the task of finding an appropriate approach to analyze the quantum models on hyperbolic lattices, which belongs to challenging problems related to the correspondence between the anti-de Sitter space and the conformal field theory~\cite{QG}, still remains an open question of the quantum gravity. A remarkable demand for an appropriate numerical tool persists. Implementation of the Monte Carlo simulations fails due to exponential increase of the number of the lattice sites for models on hyperbolic lattices with respect to the expanding lattice size from the lattice center~\cite{MC1,MC2}. Our desire is to propose a novel and sufficiently accurate numerical algorithm, which originates from the solid state physics and inherits the typical features coming from widely accepted renormalization group approaches, especially based on the Density Matrix Renormalization Group~\cite{White,Uli1,Uli2}.

Recently, we modified the Tensor Product Variational Formulation (TPVF)~\cite{DaniskaGendiar}, which is an algorithm combining an ansatz for the ground-state in the form of the Tensor product state (TPS)~\cite{Orus} with the Corner transfer matrix renormalization group (CTMRG) scheme~\cite{Nishino}. This algorithm can be used to study quantum spin systems in the thermodynamic limit on regular hyperbolic lattices of constant negative Gaussian curvature~\cite{Sadoc}. The hyperbolic lattices are constructed by tessellation of congruent $p$-sided polygons (with the coordination number fixed to four). We applied the modified TPVF algorithm to the Euclidean square ($p = 4$) and hyperbolic pentagonal ($p = 5$) lattices in order to analyze the critical phenomena of the XY, Heisenberg and transverse field Ising model (TFIM). On the square lattice numerical inaccuracy varied from $1.2\%$ in the XY model to $3.7\%$ in TFIM at the phase transition. This observation originates in the mean-field-like behavior induced by the TPS ansatz, which, as a consequence, cannot accurately approximate the correct ground state of the TFIM on the two-dimensional Euclidean lattice, which belongs to the Ising universality class. On the contrary, since the Hausdorff dimension of the hyperbolic lattices is infinite, spin models on these lattices belong to the mean-field universality class due to short range correlations, even though the mean-field approximation of the Hamiltonian is not applied~\cite{Baxter}. We conjectured that TPVF was more suitable for models on the pentagonal hyperbolic lattice due to off-critical and weakly correlated characteristics~\cite{DaniskaGendiar}.

In this work we expand the set of hyperbolic lattices investigated by the TPVF to a series of lattices constructed from congruent $p$-sided polygons, where $p \in \{5,6, \dots, 11\}$. In analogy to our previous studies of classical spin models on these hyperbolic lattices \cite{hctmrg-Ising-5-4,hctmrg-Ising-p-4,hctmrg-Ising-3-q,hctmrg-Ising-3-qn}, we expect fast convergence of the phase transition magnetic field of the quantum TFIM as well as the ground-state energies of the quantum XY and Heisenberg models toward the asymptotic case $p \rightarrow \infty$, which represents the Bethe lattice \cite{hctmrg-Ising-p-4}. Numerical results presented in the following sections are in complete agreement with the expectations. The key feature of this work is the consequent indirect analysis of the quantum TFIM, XY, and Heisenberg models on the Bethe lattice with coordination number four, which has not been considered yet.

The article is organized as follows. In Sec.~II we define the three Hamiltonians on the respective hyperbolic lattices and give a brief description of the principles of the TPVF algorithm, which have been discussed in~\cite{DaniskaGendiar}. An accurate analysis of the numerical results is presented in Sec.~III and we summarize them in Sec.~IV. 

\section{The Model}

\begin{figure}[tb]
\hspace*{2.64cm}{\includegraphics[width=0.26\textwidth,clip]{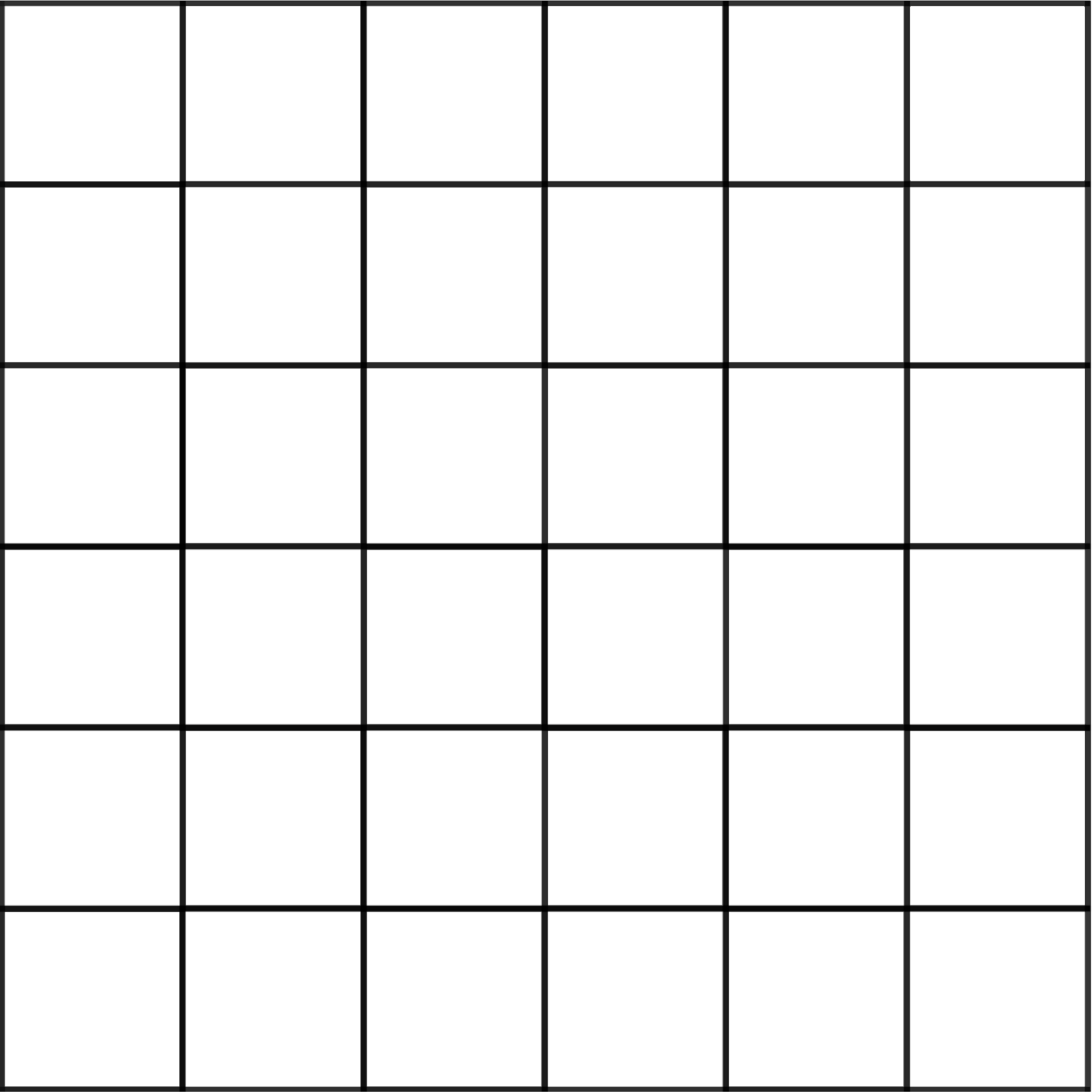}
            \includegraphics[width=0.27\textwidth,clip]{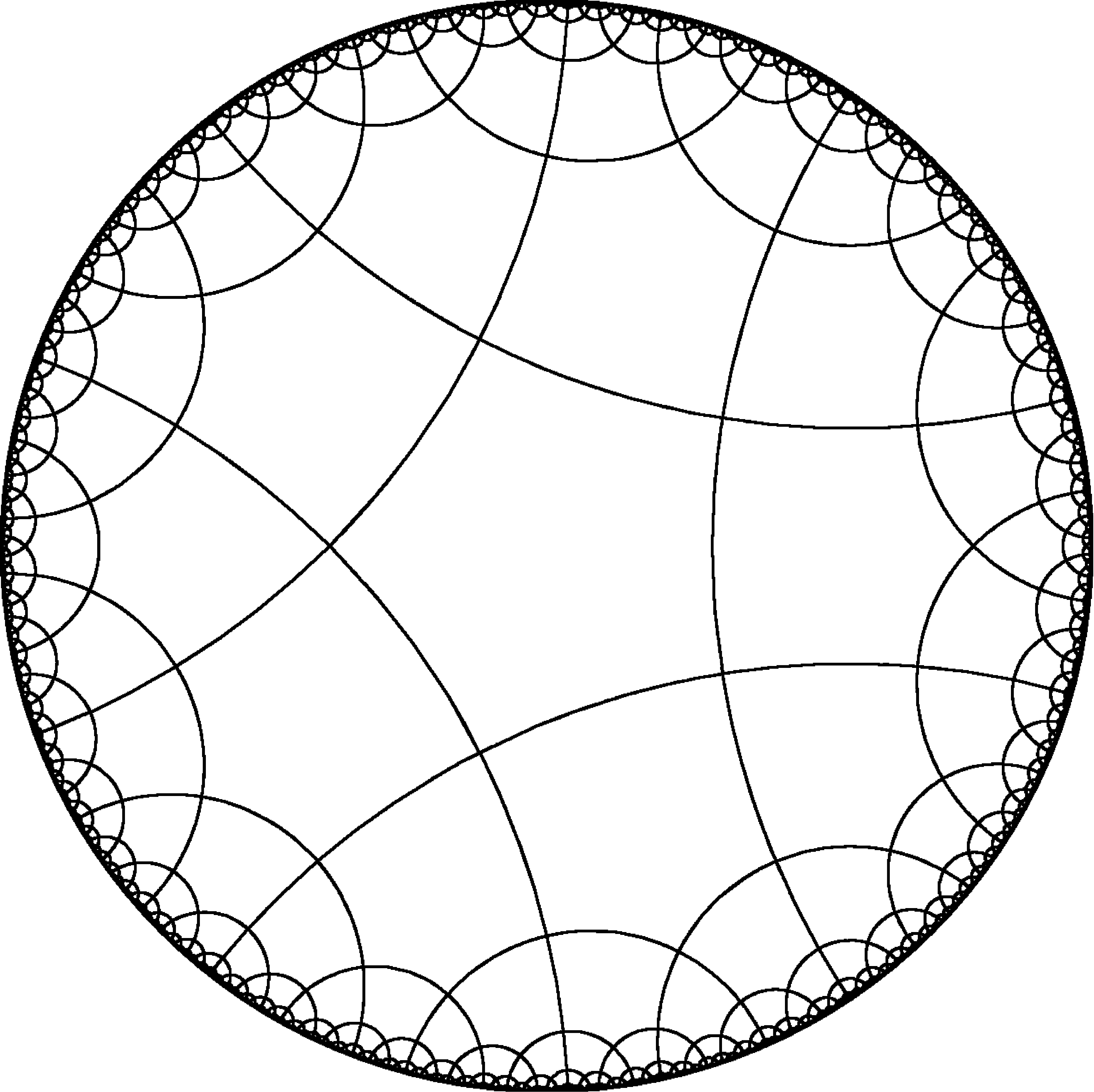}
            \includegraphics[width=0.27\textwidth,clip]{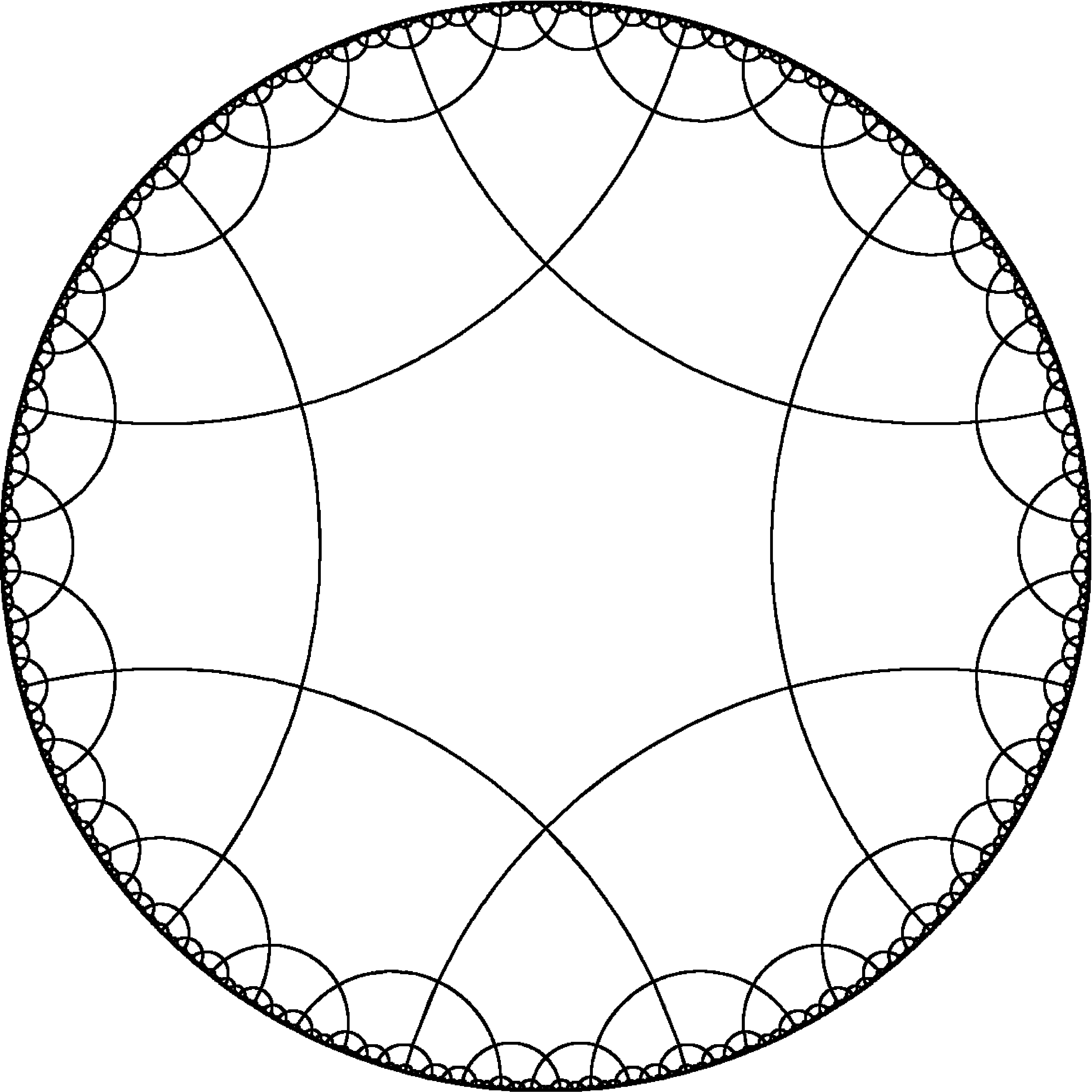}}\\
\hspace*{2.5cm}{\hfill\ \ $p=4$ \hfill\hfill $p=5$ \hfill\hfill $p=6$ \hfill{\phantom{.}}}\\

\hspace*{2.5cm}{\includegraphics[width=0.27\textwidth,clip]{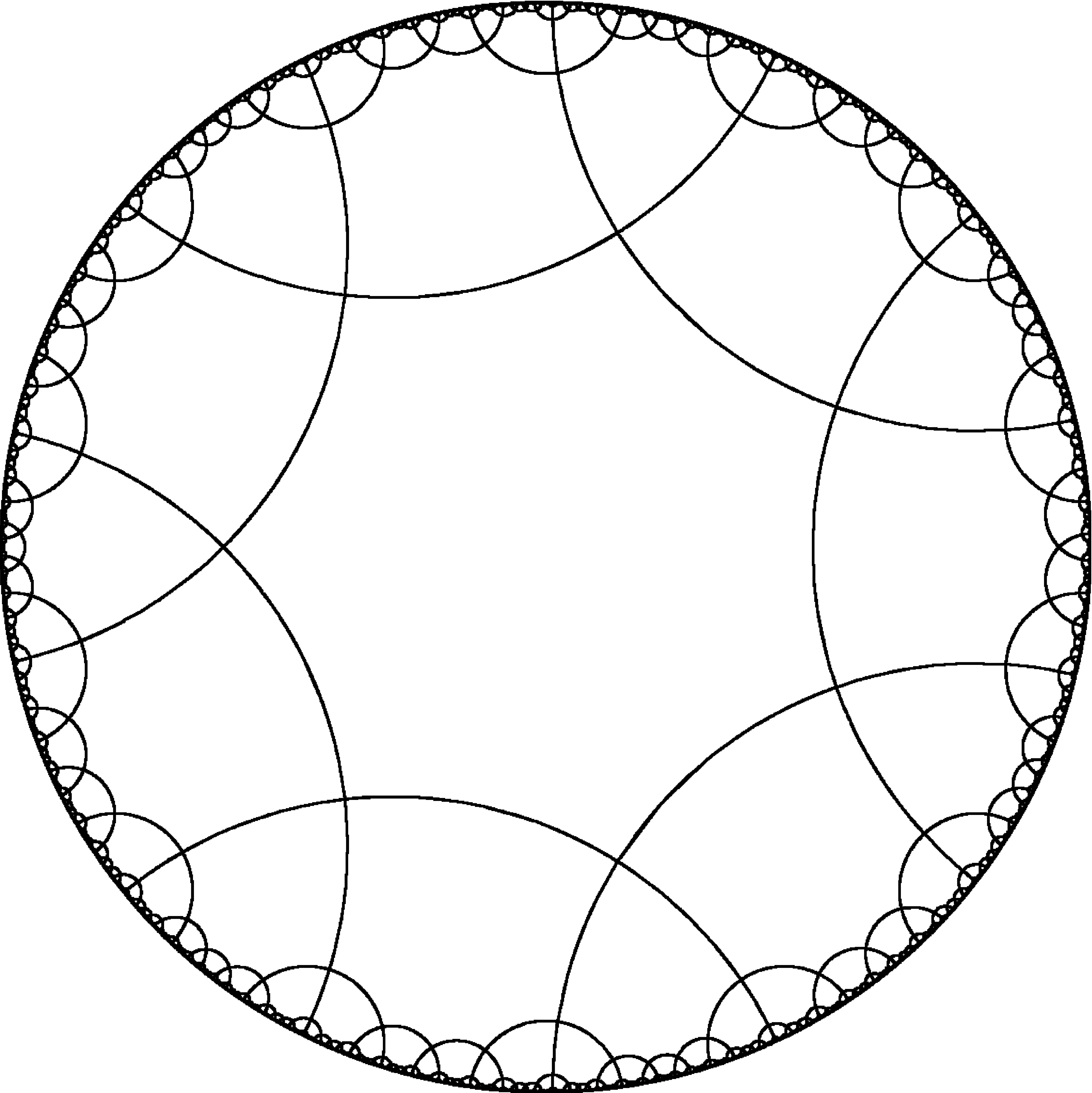}
            \includegraphics[width=0.27\textwidth,clip]{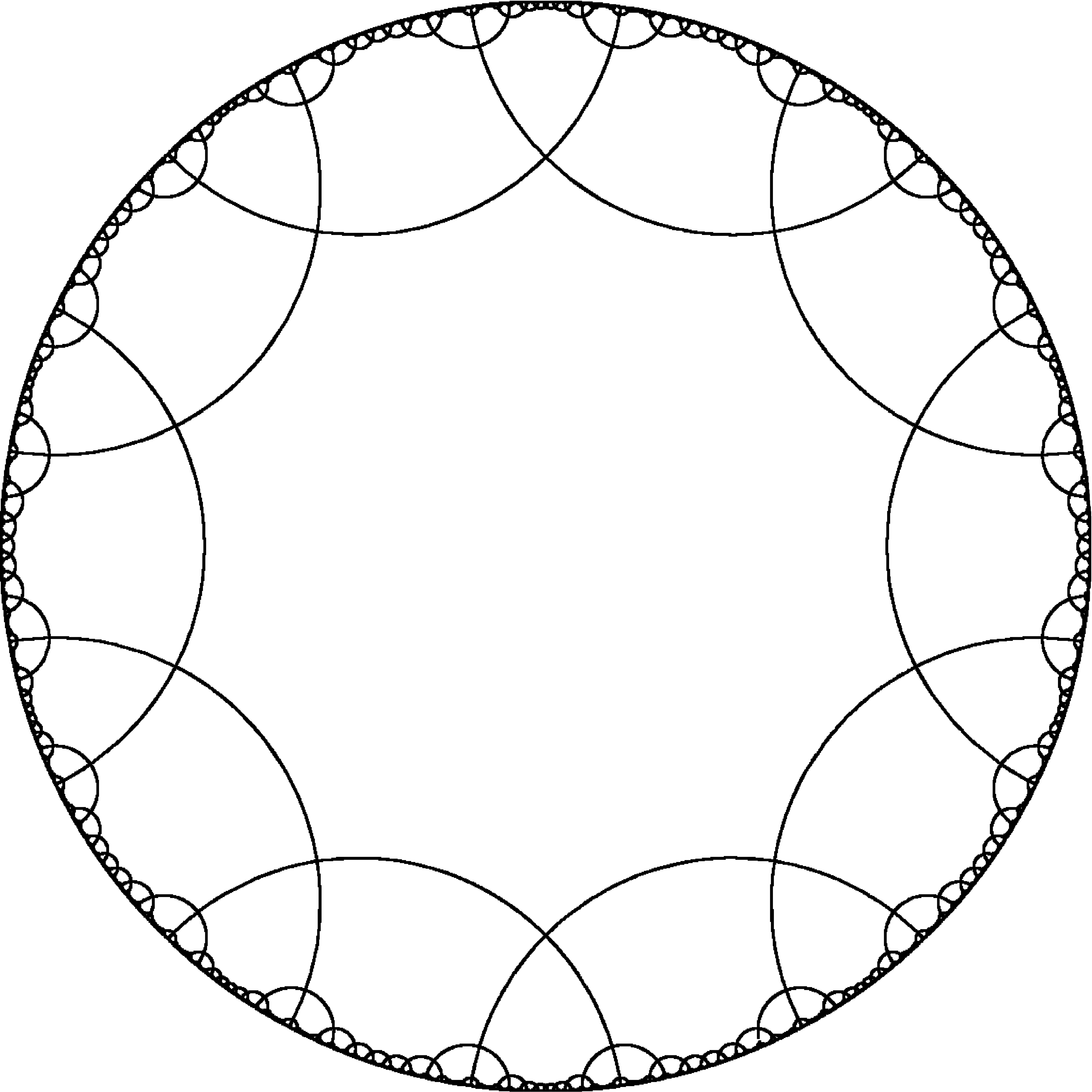}
            \includegraphics[width=0.26\textwidth,clip]{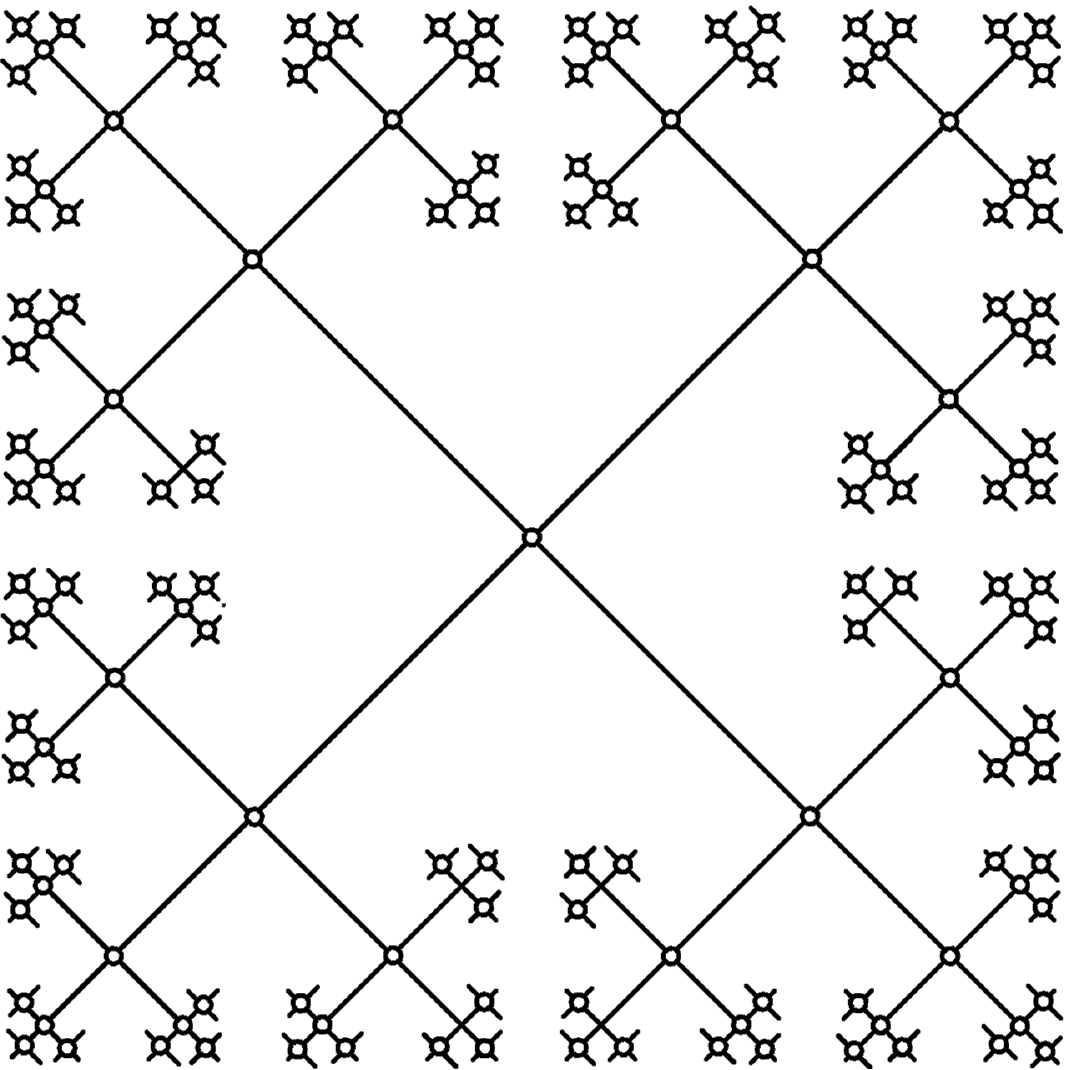}}

\hspace*{2.5cm}{\hfill\ \ $p=7$ \hfill\hfill $p=10$ \hfill\hfill $p\to\infty$ \hfill{\phantom{.}}}
\caption{Graphical representation of the lattices with the fixed coordination number equal to four indexed by the lattice parameter $p$. The hyperbolic lattices ($p=5,6,7$, and $10$) are depicted in the Poincar\'{e} disk representation, which maps the infinite-sized hyperbolic lattices onto the unitary circle, which leads to the deformation of the uniform and regular polygons toward the circle boundary.}
\label{Fig1}
\end{figure}

We study the ground-state properties and the phase transition of the quantum TFIM, XY, and Heisenberg models in the thermodynamic limit on a series of hyperbolic lattices. Each hyperbolic lattice is made from equivalent congruent $p$-sided polygons. The polygon vertices coincide with the lattice spin sites, where a single spin is positioned. Each spin site has four nearest-neighboring spin sites, which is commonly referred to as the coordination number equal to four. We investigate the three models on a set of regular hyperbolic lattices of infinite size with the lattice parameter $p \in \{5,6,\dots, 11\}$. Apart from the set, we include two additional cases: $p=4$ being the Euclidean square lattice and the asymptotic case $p\to\infty$, which is associated to the Bethe lattice. Figure~\ref{Fig1} depicts the typical structure of the lattices. The square lattice serves as a reference lattice. The three spin models and the TPVF algorithm have been described in detail in~\cite{DaniskaGendiar}, and we focus only on the substantial aspects of the models on the hyperbolic lattices in the following.

In general, the Hamiltonian ${\cal H}$ of any of the three models can be written as a sum of local Hamiltonians $G_{k}^{(p)}$ of the $p$-sided polygonal shape, in particular,  
\begin{equation}
{\cal H} = \sum\limits_{{\langle k \rangle}_p}^{~} G_{k}^{(p)}\, ,
\label{Hm1}
\end{equation}
where $k$ labels the polygons and the sum runs over the set of all indices of the lattice polygons ${\langle k \rangle}_p$. The local Hamiltonian takes the form
\begin{eqnarray}
\nonumber
G_{k}^{(p)} & = & -\frac{1}{2}\sum\limits_{i=1}^{p} \bigg[
      J_{xy} \left( S_{k_i}^{x} S_{k_{i+1}}^{x} + S_{k_i}^{y} S_{k_{i+1}}^{y} \right) \\
      & + & J_{z} S_{k_i}^{z} S_{k_{i+1}}^{z}
      + \frac{h}{4} \left( S_{k_i}^{x} + S_{k_{i+1}}^{x}\right) \bigg] \, ,
\label{Hm2}
\end{eqnarray}
where $k_1,k_2,\dots,k_p$ label the spin positions on the $k^{\rm th}$ $p$-sided polygon (noticing that $k_{p+1} \equiv k_1$), and $S_{k_i}^x$, $S_{k_i}^y$, $S_{k_i}^z$ denote the corresponding Pauli spin-$\frac{1}{2}$ operators. We consider constant nearest-neighbor couplings $J_{xy}, J_{z}$ and a uniform external magnetic field $h$. By setting $J_{xy} = 0$ and $J_{z} = 1$ we obtain the TFIM at the transverse magnetic field $h$, whereas the choice $J_{xy} = 1,J_{z} = h = 0$ gives the XY model and $J_{xy} = - J_{z} = 1, h = 0$ the Heisenberg model ~\cite{DaniskaGendiar}. 

Our task is to calculate an approximate ground-state of the system in the thermodynamic limit in the product form
\begin{equation}
      | \Psi_p\rangle = \lim\limits_{N\to\infty}
      \sum\limits_{\sigma_1^{~}\sigma_2^{~}\cdots\sigma_{N}^{~}}^{~}
      \prod\limits_{{\langle k \rangle}_p}^{~} W_p(\{\sigma_k\}) 
      |\sigma_1^{~}\sigma_2^{~}\cdots\sigma_N^{~}\rangle \, ,
\label{Psi}
\end{equation}
where $N$ stands for the total number of the lattice spins. The basis $\sigma_j$ for $j=1, \dots, N$ denotes a binary state, for which we use the arrow notation $\dn$ or $\up$ in the following. The summation runs over the $2^N$ base spin states $|\sigma_1^{~}\sigma_2^{~}\cdots\sigma_N^{~}\rangle$, and $W_p(\{ \sigma_k \})$ are the elements of the $p$-rank tensor $W_p$ depending on $p$ spins $\sigma_{k_1}, \dots, \sigma_{k_p}$ on the $k^{th}$ lattice polygon. The symbol $\{\sigma_k\}$ stands for one of the $2^p$ base configurations of a multi-spin variable representing the group of spins $\sigma_{k_1}, \dots, \sigma_{k_p}$. All the tensors $W_p$ are considered to be identical, therefore, the set of  $2^p$ tensor elements $W_p(\{\sigma\})$, where the subscript $k$ has been omitted due to the uniformity of the tensors $W_p$, uniquely describes the state $ |\Psi_p\rangle$, i.e. $|\Psi_p\rangle = |\Psi_p[W_p(\{\sigma\})]\rangle$. 
 
We regard $|\Psi_p^{*}\rangle$ as the best approximation of the ground-state within the class of TPS $|\Psi_p\rangle$, if the minimum of the energy normalized per bond,
\begin{equation}
{ E}_0^{(p)} \equiv \min\limits_{\Psi_p}\lim\limits_{N_b\to\infty}\frac{1}{N_b}
            \frac{\langle \Psi_p | {\cal H} | \Psi_p \rangle}
                 {\langle \Psi_p            | \Psi_p \rangle}                
                  \, ,
\label{Eg}
\end{equation}
is obtained for $|\Psi_p^{*}\rangle$. Here, $N_b$ denotes the total number of bonds in the system. The energy ${E}_0^{(p)}$, due to its variational origin, serves as an upper bound of the true ground-state energy per bond ${\cal E}_{0}^{(p)}$.

Since the structure of every local Hamiltonian $G_{k}^{(p)}$ does not depend on $k$ (we investigate the system in the thermodynamic limit), the variational problem in~\eref{Eg} is equivalent to minimization of the local energy per bond of an arbitrary polygon in the lattice center (in order to avoid boundary effects)
\begin{equation}
{E}_0^{(p)} = \min\limits_{\Psi_p}
            \frac{2}{p}\frac{\langle \Psi_p | {G_\ell^{(p)}} | \Psi_p \rangle}
                 {\langle \Psi_p            | \Psi_p \rangle} \, ,
\label{Egbond}
\end{equation}
where $\ell$ is the index of the selected polygon and the normalization factor $2/p$ reflects the fact that the $p$ bonds of each polygon are shared with its neighbors. Moreover, if we utilize the tensor product structure of the state $|\Psi_p\rangle$, we can express the denominator $\langle \Psi_p | \Psi_p \rangle \equiv {\cal D}(W_p(\{\sigma\}))$ and the numerator $\langle \Psi_p | {G_\ell^{(p)}} | \Psi_p \rangle \equiv {\cal N}(W_p(\{\sigma\}))$ as functions of the tensor elements $W_p(\{\sigma\})$ only. Consequently, our variational problem transforms onto a multi-dimensional minimization over $2^p$ tensor elements $W_p(\{\sigma\})$
\begin{equation}
{E}_0^{(p)} =  \min\limits_{W_p(\{\sigma\})} 
               \frac{2}{p}\frac{{\cal N}(W_p(\{\sigma\}))}{{\cal D}(W_p(\{\sigma\}))} \, .
\label{e0}
\end{equation}

Furthermore, symmetries of the local Hamiltonian $G_{\ell}^{(p)}$ may significantly reduce the dimension of the problem. Rotational and spin-ordering symmetries are present in all the three spin models. As a typical example, let us consider a hexagonal lattice ($p=6$) and its particular base configuration of spins on the lattice polygon $\{\sigma^{*}\} = \{\up\dn\up\up\dn\dn\}$. Rotational symmetry requires that the tensor elements corresponding to the set of configurations $\{\dn\up\dn\up\up\dn\}$, $\{\dn\dn\up\dn\up\up\}$, $\{\up\dn\dn\up\dn\up\}$,  $\{\up\up\dn\dn\up\dn\}$, $\{\dn\up\up\dn\dn\up\}$ are identical to $W_{p=6}(\{\sigma^{*}\})$. Next, let us consider a spin-ordering operation, which reverses the order of the polygon spins. In particular, if the spins are labelled clockwise, the operation reorders them in the anti-clockwise direction. It means that the configuration $\{\up\dn\up\up\dn\dn\}$ is equivalent to $\{\dn\dn\up\up\dn\up\}$ by the spin-ordering symmetry and to all the rotations of the latter configuration ($\{\up\dn\dn\up\up\dn\}$, $\{\dn\up\dn\dn\up\up\}$, $\{\up\dn\up\dn\dn\up\}$, $\{\up\up\dn\up\dn\dn\}$, $\{\dn\up\up\dn\up\dn\}$) by a composition of the spin-ordering and the rotational symmetry. As a result, the 12 tensor elements $W_6(\{\sigma\})$ corresponding to the configuration $\{\sigma^{*}\}$ and its 11 equivalent configurations are represented by a single variational parameter, as they share the same value.

By performing a similar analysis on the set of all $2^p$ configurations $\{\sigma\}$ we can factorize it into $N_{\rm Ising}^{(p)}$ classes of equivalence with representatives $\theta_j$, where $j \in \{1, \dots, N_{\rm Ising}^{(p)}\}$~\cite{DaniskaGendiar}. Thus, in case of a system with the rotational and the spin-ordering symmetry (as in the TFIM), there are only $N_{\rm Ising}^{(p)}$ free variational parameters $W_p(\theta_j)$ within the set of $2^p$ tensor elements $W_p(\{\sigma\})$. If there is no preferred spin alignment in the system (such as in the XY model, the Heisenberg model, as well as in the TFIM at and above the phase transition magnetic field), the spin-inversion symmetry appears. For instance, if $p=4$, the configuration $\{\up\up\up\dn\}$ is equivalent to $\{\dn\dn\dn\up\}$, which is obtained by flipping each spin. Such an additional symmetry results in consequent reduction of the set of the free variational parameters, the size of which drops to $N_{\rm Heis}^{(p)} < N_{\rm Ising}^{(p)}$. The numbers of the free variational parameters $N_{\rm Ising}^{(p)}$ and $N_{\rm Heis}^{(p)}$ with respect to the lattice parameter $p$ are summarized in table~\ref{Tab1}. In addition, one more variational parameter can be eliminated from each set of the free variational parameters by setting it to $1$, being the normalization condition in $W_p(\{\sigma\})$ and $| \Psi_p \rangle$, consequently.

\begin{table}[tb]
\caption{The numbers of the free variational parameters $N_{\rm Heis}^{(p)}$ (for the XY and the Heisenberg models) and $N_{\rm Ising}^{(p)}$ (for the TFIM) including the normalization parameter.}
\begin{center}
\begin{tabular}{| c || r | r | r | r | r | r | r | r | }
\hline
$p$ & $4$ & $5$ & $6$ & $7$ & $8$ & $9$ & $10$ & $11$  \\
\hline
\hline
$N_{\rm Heis}^{(p)}$ & $4$ & $4$ & $8$ & $9$& $18$ & $23$ & $44$ & $63$ \\[0.04cm]
\hline
$N_{\rm Ising}^{(p)}$ & {\phantom{00}}$6$ & {\phantom{00}}$8$ & {\phantom{0}}$13$ & {\phantom{0}}$18$&
{\phantom{0}}$30$ & {\phantom{0}}$46$ & {\phantom{0}}$78$ & $126$ \\[0.04cm]
\hline
\end{tabular}
\end{center}
\label{Tab1}
\end{table}

The free variational parameters $W_p(\theta_j)$ are optimized numerically by TPVF~\cite{DaniskaGendiar}. It is based on the fact that the product structure of the state $|\Psi_p\rangle$ allows to calculate the numerator ${\cal N}(W_p(\{\sigma\}))$ and the denominator ${\cal D}(W_p(\{\sigma\}))$ in~\eref{e0} for the given set of the tensor elements $W_p(\{\sigma\})$ by an appropriate modification of the CTMRG algorithm. Having applied the CTMRG as the effective and accurate numerical tool for calculation of the ratio in~\eref{e0}, a multi-dimensional minimizer is used for optimizing the variational parameters $W_p(\theta_j)$~\cite{gsl1,gsl2,NM}.

\section{Numerical results}
\subsection{XY and Heisenberg models}

We study the XY and the Heisenberg models at zero magnetic field, where these models are known to be critical in the Euclidean space. Therefore, there is no preferred direction (the spin alignment) in the system on the Euclidean lattice at $h \geq 0$, and the spin-inversion symmetry is present. We expect that the models on hyperbolic lattices also exhibit the spin-inversion symmetry. It enables to reduce the number of the free variational parameters $W_p(\theta_j)$ within the TPVF minimization part down to $N_{\rm Heis}^{(p)}$ as listed in table~\ref{Tab1}. Despite the significant reduction, the number of the free parameters $N_{\rm Heis}^{(p)}$ still grows fast with respect to the increasing lattice parameter $p$. The computational time of the minimization algorithm is significantly prolonged due to (at least) linear dependence on the increasing number of the free variational parameters. Also, the algorithm may possibly be trapped in a local energy minimum and thus a series of initial conditions has to be tested in order to obtain the global energy minimum (or, at least, a sufficiently good approximation of it). For all these reasons, the calculations were stopped at $p=11$ with respect to the constraints of our computational resources and time.

\tabcolsep=3pt
\begin{table}[tb]
\caption{The ground-state energies per bond $E_0^{(p)}$ listed with respect to $p$ for the Heisenberg and XY models. The number of block spin states~\cite{White,Nishino} kept was $m=20$ for $4\leq p \leq 10$ and $m=10$ for $p=11$. The asymptotic estimate of $E_0^{(\infty)}$ corresponds to the model on the Bethe lattice.}
\begin{center}
\begin{tabular}{| c | c | c | }
\hline
\multirow{2}{*}{$p$} & \multicolumn{2}{ c |}{$E_0^{(p)}$}  \\ \cline{2-3}
  & XY & Heisenberg  \\ \hline
$4$      & $-1.08456618$ & $-1.3089136$ \\ \hline
$5$      & $-1.08151200$ & $-1.2912704$ \\ \hline
$6$      & $-1.08097046$ & $-1.2925639$ \\ \hline
$7$      & $-1.08086301$ & $-1.2918936$ \\ \hline
$8$      & $-1.08084068$ & $-1.2919769$ \\ \hline
$9$      & $-1.08083585$ & $-1.2919403$ \\ \hline
$10$     & $-1.08083478$ & $-1.2919460$ \\ \hline
$11$     & $-1.08083453$ & $-1.2919437$ \\ \hline
$\infty$ & $-1.08083446$ & $-1.291944${\phantom{0}} \\ \hline
\end{tabular}
\end{center}
\label{Tab2}
\end{table}
\tabcolsep=2pt

The ground-state energies $E_0^{(p)}$ obtained by the TPVF algorithm for both the XY and the Heisenberg models are summarized in table~\ref{Tab2}. The energies $E_0^{(p)}$ remained identical even if the larger set of $N_{\rm Ising}^{(p)}$ free variational parameters $W_p(\theta_j)$ in TPVF was used, whereby the optimal values of the parameters $W^{*}_p(\theta_j)$ coupled by spin-inversion symmetry were equal. These results witness the spin-inversion symmetry of the models on hyperbolic lattices.  Recall that $E_0^{(p)}$ represents only an upper estimate of the true ground-state energy ${\cal E}_0^{(p)}$. We have shown that the energies $E_0^{(4)}$ of the referencing Euclidean square lattice calculated by TPVF were higher if compared to the Monte Carlo simulations (the relative errors for the XY and the Heisenberg models, respectively, are $1.2\%$ and $2.2\%$)~\cite{DaniskaGendiar}. This observation can be explained by suppression of the quantum long-range correlations induced by the TPS approximation of the low-dimensional uniform tensors $W_4$, which cannot correctly reproduce the divergence of the correlation length in the models on the square lattice. On the other hand, any quantum spin model on hyperbolic lattice belongs to the mean-field universality class, because the hyperbolic lattices exhibit the infinite Hausdorff dimension, which significantly exceeds the critical lattice dimension $D_c=3$~\cite{Baxter}. Because of the mean-field-like character of the TPS approximation, the TPVF algorithm is expected to be more accurate whenever a hyperbolic lattice geometry is considered~\cite{DaniskaGendiar,hctmrg-Ising-p-4}.

 \begin{figure}[tb]
\centerline{\includegraphics[width=0.7\textwidth,clip]{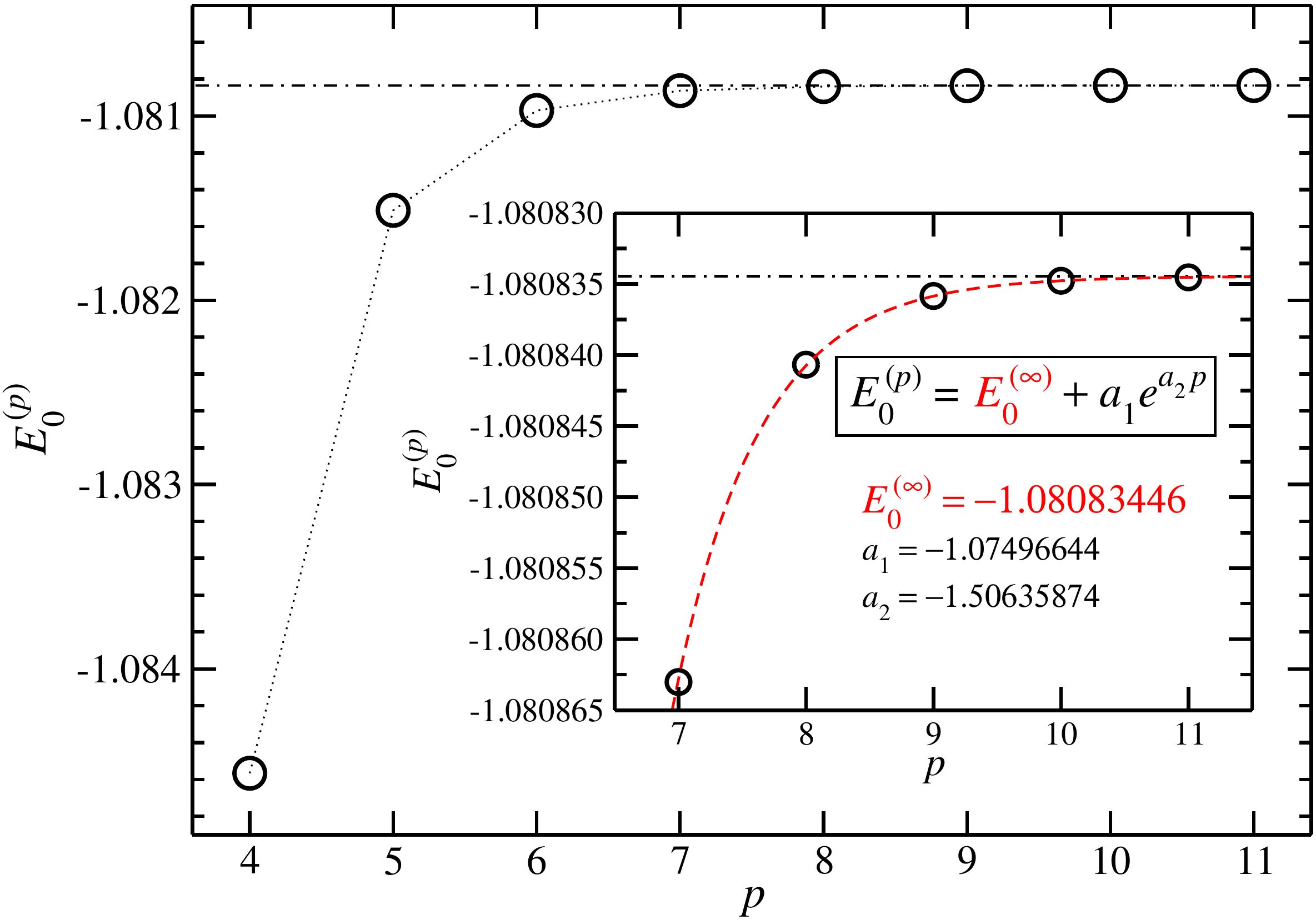}}
\caption{The ground-state energy $E_0^{(p)}$ of the XY model with respect to the lattice parameter $p \in \{4,5,\dots, 11\}$. The inset shows the zoomed-in energy including the details of the fitting function.}
\label{Fig2}
\end{figure}

\begin{figure}[tb]
\centerline{\includegraphics[width=0.7\textwidth,clip]{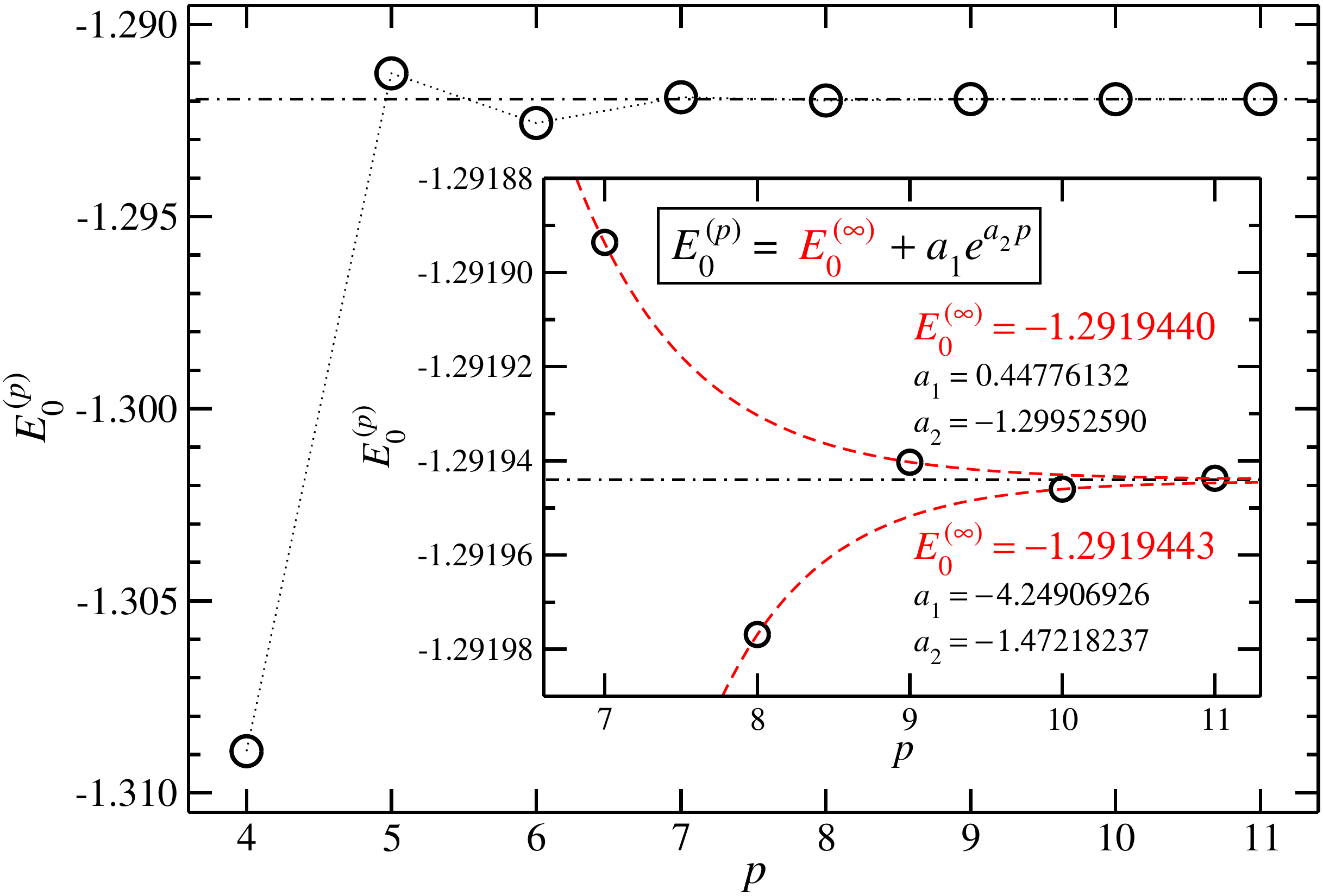}}
\caption{The ground-state energy $E_0^{(p)}$ of the Heisenberg model with respect to the lattice parameter $p \in \{4,5,\dots, 11\}$. The fitting function parameters are shown in the inset.}
\label{Fig3}
\end{figure}

Figure \ref{Fig2} illustrates a monotonically increasing and rapidly saturating curve of the energy $E_0^{(p)}$ for the XY model with respect to the lattice parameter $p$. The inset depicts the tail of the curve in detail together with an exponential fit applied to the five energies $E_0^{(7)}, \dots, E_0^{(11)}$. The case $p=4$, where the TPVF algorithm is not sufficiently accurate for the reasons mentioned above, was excluded from the extrapolation analysis. The fitting function is proposed in the form
 \begin{equation}
 E_0^{(p)} = E_0^{(\infty)} +a_1 \exp(a_2 p ) \, ,
 \label{fitXY}
\end{equation}
 where $E_0^{(\infty)}$, $a_1$, and $a_2$ are the fitting parameters, which were determined in the following way. First we defined a function $f(E)$, which returns the residual sum of squares ($RSS$) of the linear regression $\ln\vert E-E_0^{(p)}\vert = \ln\vert a_1\vert +a_2 p $. Then, $E_0^{(\infty)}$ was chosen as the argument, which minimizes the function $f(E)$. The corresponding linear regression $\ln\vert E_0^{(\infty)}-E_0^{(p)}\vert = \ln\vert a_1\vert + a_2 p $ specifies the parameters $a_1$ and $a_2$. If considering another way, $E_0^{(\infty)}$ is such a value that the curve $\ln\vert E_0^{(\infty)}-E_0^{(p)}\vert$ is as close as possible to a line, where the closeness is measured by the $RSS$. Thus obtained parameters $E_0^{(\infty)}$, $a_1$, and $a_2$ are listed in the inset of figure~\ref{Fig2}, where the dot-dashed line represents the estimate of the ground-state energy per bond of the quantum XY model on the Bethe lattice $E_0^{(\infty)}=\lim\limits_{p \rightarrow \infty} E_0^{(p)}=-1.08083446$.
 
Analogously, the ground-state energies $E_0^{(p)}$ of the Heisenberg model are plotted in figure~\ref{Fig3}. Again, rapid convergence of the energy to the asymptotic values is obvious from the data. Although we have not clarified the physical origin of the non-monotonic convergence (saw-like pattern) of $E_0^{(p)}$ yet, a detailed analysis indicates that the exponential fitting function in~\eref{fitXY} can successfully describe the data, if applied separately onto two sets: those with even $p \in \{6, 8, 10\}$ (the lower branch shown in the inset) and the odd $p \in \{5, 7, 9, 11\}$ (the upper branch). The fitting parameters of the two regressions are listed in the inset of figure~\ref{Fig3}. The lower and the upper branches yield the energies $E_0^{(\infty)}$ $-1.2919443$ and $E_0^{(\infty)}-1.2919440$, respectively. With respect to an independent application of additional analogous fits, we found $E_0^{(\infty)}=-1.291944$ (all the digits are valid) to be considered as the correct estimate of the ground-state energy per bond of the Heisenberg model on the Bethe lattice.        

We have not found any theoretical reasoning for the exponential convergence of the ground-state energies $E_0^{(p)}$ yet. However, if a power-law fitting function was applied instead, we obtained a less accurate fitting and greater $RSS$.

\subsection{Transverse field Ising model}

The TFIM undergoes a quantum phase transition at a nonzero magnetic field $h_t^{(p)}>0$, where we explicitly emphasize its dependence on the lattice geometry. The nonzero spontaneous magnetization in the ordered phase at $h<h_t^{(p)}$ breaks the spin-inversion symmetry, which results in approximately twice larger set of the free variational parameters $N_{\rm Ising}^{(p)}$ in the TPVF algorithm if compared to $N_{\rm Heis}^{(p)}$ in the XY and Heisenberg models, cf.~table~\ref{Tab1}. The computational time for a particular fixed field $h$ is, therefore, significantly prolonged. Moreover, in order to screen the vicinity of the phase transition field $h_t^{(p)}$, multiple calculations for a sequence of magnetic fields $h$ had to be performed. As a consequence, in order to restrict the total computational time, we have analyzed the TFIM on the hyperbolic lattices up to $p=10$ only. (Notice that the number of block spins states kept was $m=20$ for $p \in \{4,5, \dots, 8\}$, and only $m=4$ for $p \in \{9,10\}$, which was sufficient due to exponentially weak correlations caused by the hyperbolic lattice geometry~\cite{hctmrg-Ising-3-q}; any further increase of the states kept $m$ has not improved the numerical calculations significantly).

\begin{figure}[tb]
\centerline{\includegraphics[width=0.7\textwidth,clip]{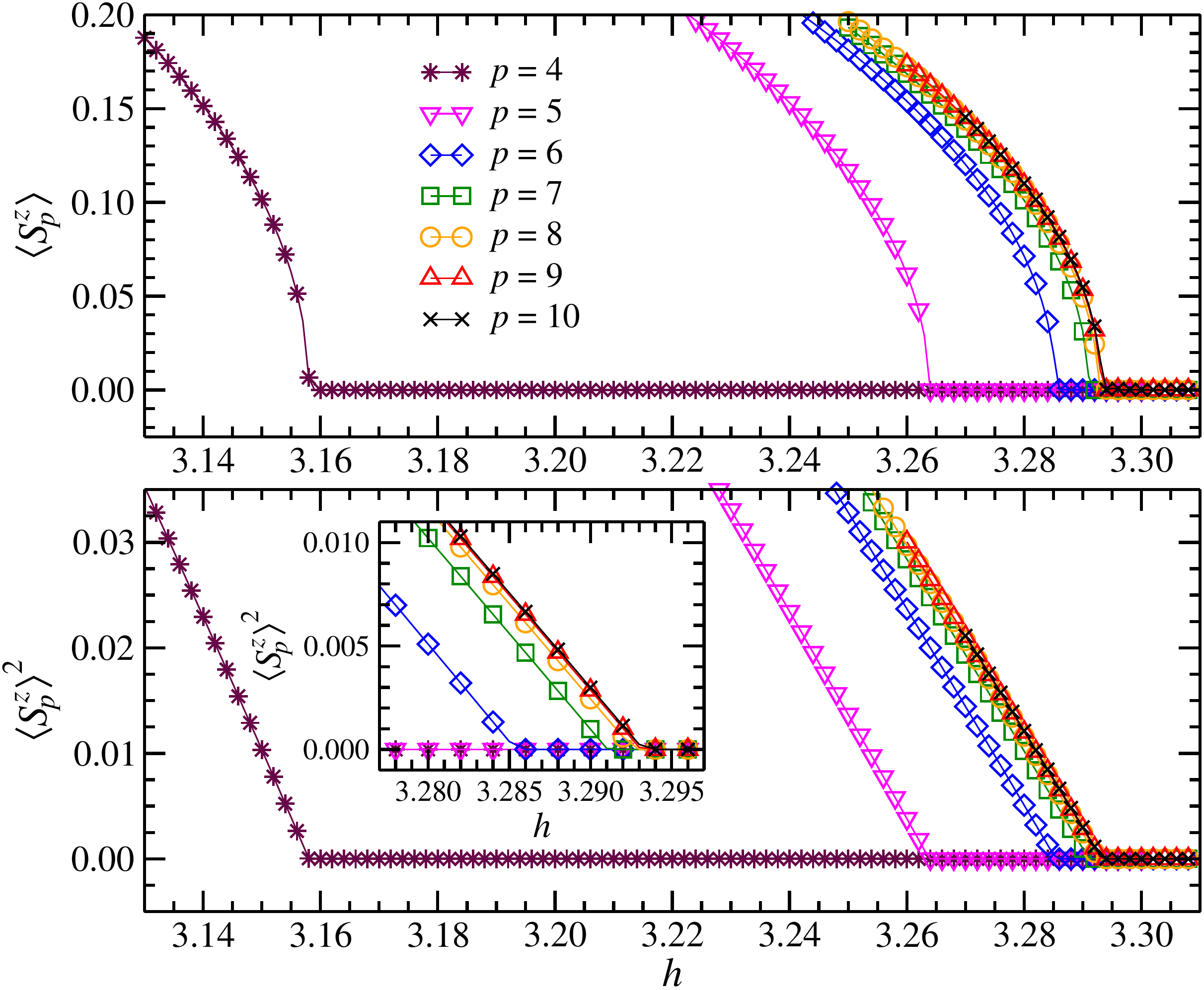}}
\caption{The spontaneous magnetization $\langle S_p^{z}\rangle$ (the upper
graph) and its square ${\langle S_p^{z}\rangle}^{2}$ (the lower graph) in the vicinity of the phase transitions with respect to the magnetic field $h$ for $p \in \{4,5, \dots, 10\}$. The inset shows the detailed zoomed-in behavior for higher values of $p$.}
\label{Fig4}
\end{figure}

We have analyzed the phase transition of the TFIM by the expectation value of the spontaneous magnetization $\langle S_p^{z} \rangle$ as well as by the magnetic susceptibility $\chi_p$. Solving the minimization problem in~\eref{e0}, we received the optimal tensor elements $W^*_p(\{\sigma\})$, which uniquely define the approximative ground state $|\Psi_p^*\rangle$ via~\eref{Psi}. Once $|\Psi_p^*\rangle$ has been constructed, we evaluated the spontaneous magnetization   
\begin{equation}
\langle S_p^{z} \rangle = \frac{\langle \Psi^*_p | S_{c}^{z} |\Psi^*_p \rangle}
                                    {\langle \Psi^*_p |                   \Psi^*_p \rangle} \, ,
\label{sz}
\end{equation}
where ${c}$ labels an arbitrary spin in the central polygon of the lattice in order to suppress boundary effects. The resulting dependence of the magnetization $\langle S_p^{z}\rangle$ with respect
to the magnetic field $h$ near the phase transition field $h_t^{(p)}$ is plotted in the upper graph of figure~\ref{Fig4}. The quantum phase transition of the TFIM is characterized by a non-analytic behavior of the magnetization curve, when $\langle S_p^{z}\rangle \to 0$ if approaching the phase transition field $h \to h_t^{(p)}$ from the ordered phase ($h < h_t^{(p)}$). The phase transition exponent $\beta_p$, which depends on the lattice geometry, describes the singularity through the scaling relation in the ordered phase
\begin{equation}
\langle S_p^z (h) \rangle \propto {\left(h_t^{(p)} - h \right)}^{\beta_p} \, .
\label{powerlaw}
\end{equation}
Figure~\ref{Fig4} (the lower graph) shows the squared transversal magnetization $ {\langle S_p^{z}\rangle}^{2}$, where we point out the linearity of the squared magnetization if approaching the phase transition field $h_t^{(p)}$. Such a dependence confirms the mean-field exponent $\beta_p=\frac{1}{2}$ regardless of the lattice parameter $p$, which results in the mean-field-like behavior of the TFIM if approaching the phase transition.

\begin{figure}[tb]
\centerline{\includegraphics[width=0.7\textwidth,clip]{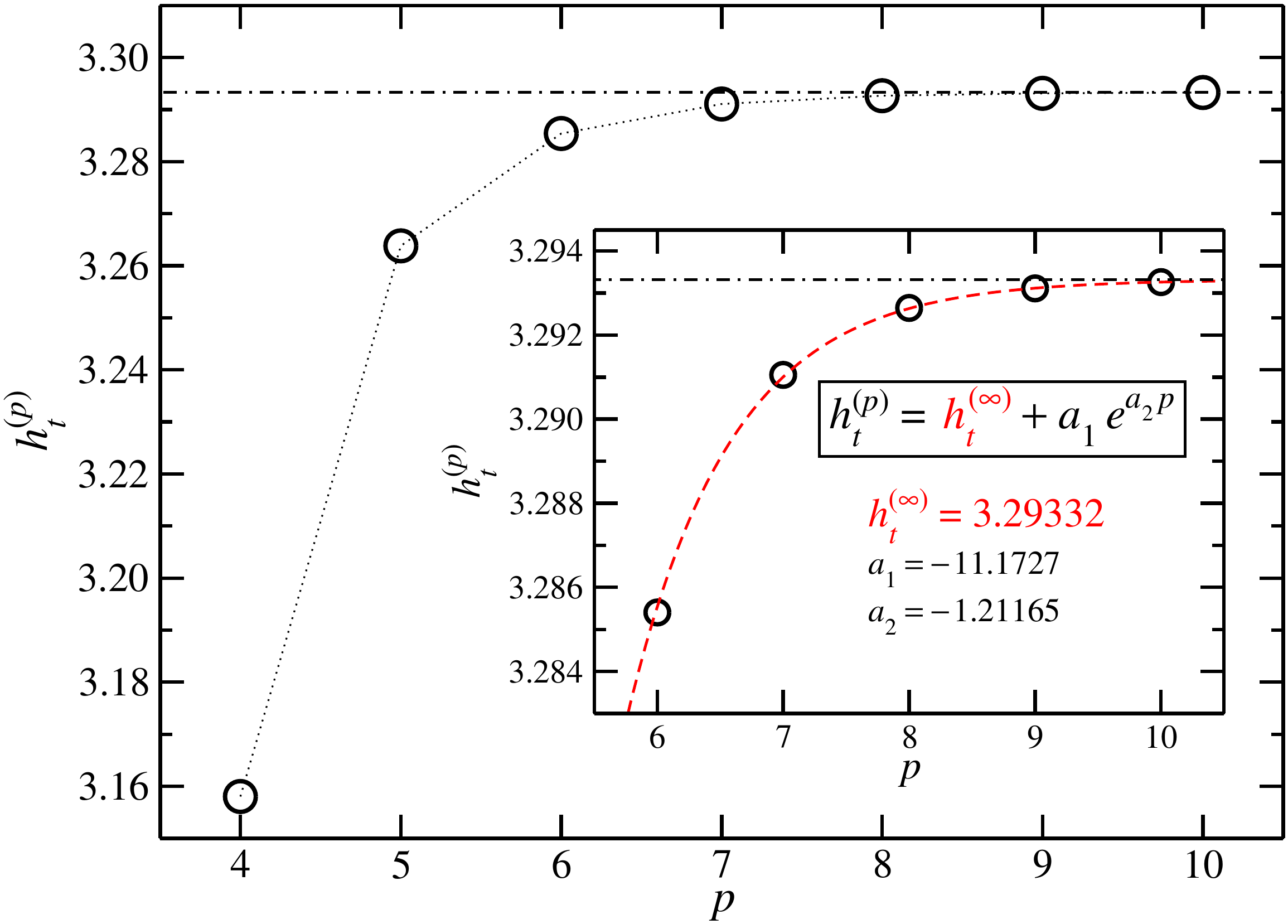}}
\caption{The phase transition field $h_t^{(p)}$ of the TFIM with respect
to the lattice parameter $p$. The horizontal dot-dashed line represents the estimated
asymptotic value $h_t^{(\infty)}=3.29332$.}
\label{Fig5}
\end{figure}

The phase transition fields $h_t^{(p)}$, calculated according to the method described in \cite{DaniskaGendiar}, are summarized in table~\ref{Tab3} together with their errors $\Delta^{(p)}$. Notice that $\Delta^{(p)}$ represents only the error of the method providing that the calculated magnetization $\langle S_p^{z}\rangle$ is considered accurate. The data are graphically plotted in figure~\ref{Fig5}, whereas the error bars are too small to be displayed. Using an analogous exponential fitting function applied on the critical magnetic fields $h_t^{(p)}$ for $p \in \{6, \dots, 10\}$, (cf.~\eref{fitXY}), we calculated the asymptotic phase transition field of the TFIM on the Bethe lattice $h_t^{(\infty)}=3.29332$ as listed in Tab.~\ref{Tab3}.

\tabcolsep=4pt
\begin{table}[tb]
\caption{The phase transition fields $h_t^{(p)}$ of the TFIM including the estimated errors $\Delta^{(p)}$ with respect to the lattice parameter $p$.}
\begin{center}
\begin{tabular}{| c | c | c | c | c | }
\hline
 $p$ 		&	$4$ & $5$ & $6$ & $7$  \\ \hline
 $h_t^{(p)}$& 	$3.158034$ & $3.263825$ & $3.285405$ & $3.291055$  \\ \hline
  $\Delta^{(p)}$ 	&	$1\times10^{-6}$ & $1\times10^{-6}$ & $1\times10^{-6}$ & $1\times10^{-6}$  \\ \hline\hline
 $p$ 		&	$8$ & $9$ & $10$ & $\infty$ \\ \hline
 $h_t^{(p)}$& 	$3.292647$ & $3.293113$ & $3.293263$& $3.29332$  \\ \hline
  $\Delta^{(p)}$ 	&	$2\times10^{-6}$ & $2\times10^{-6}$ & $5\times10^{-6}$ & $1\times10^{-5}$  \\ \hline
\end{tabular}
\end{center}
\label{Tab3}
\end{table}
\tabcolsep=2pt

\begin{figure}[tb]
\centerline{\includegraphics[width=0.7\textwidth,clip]{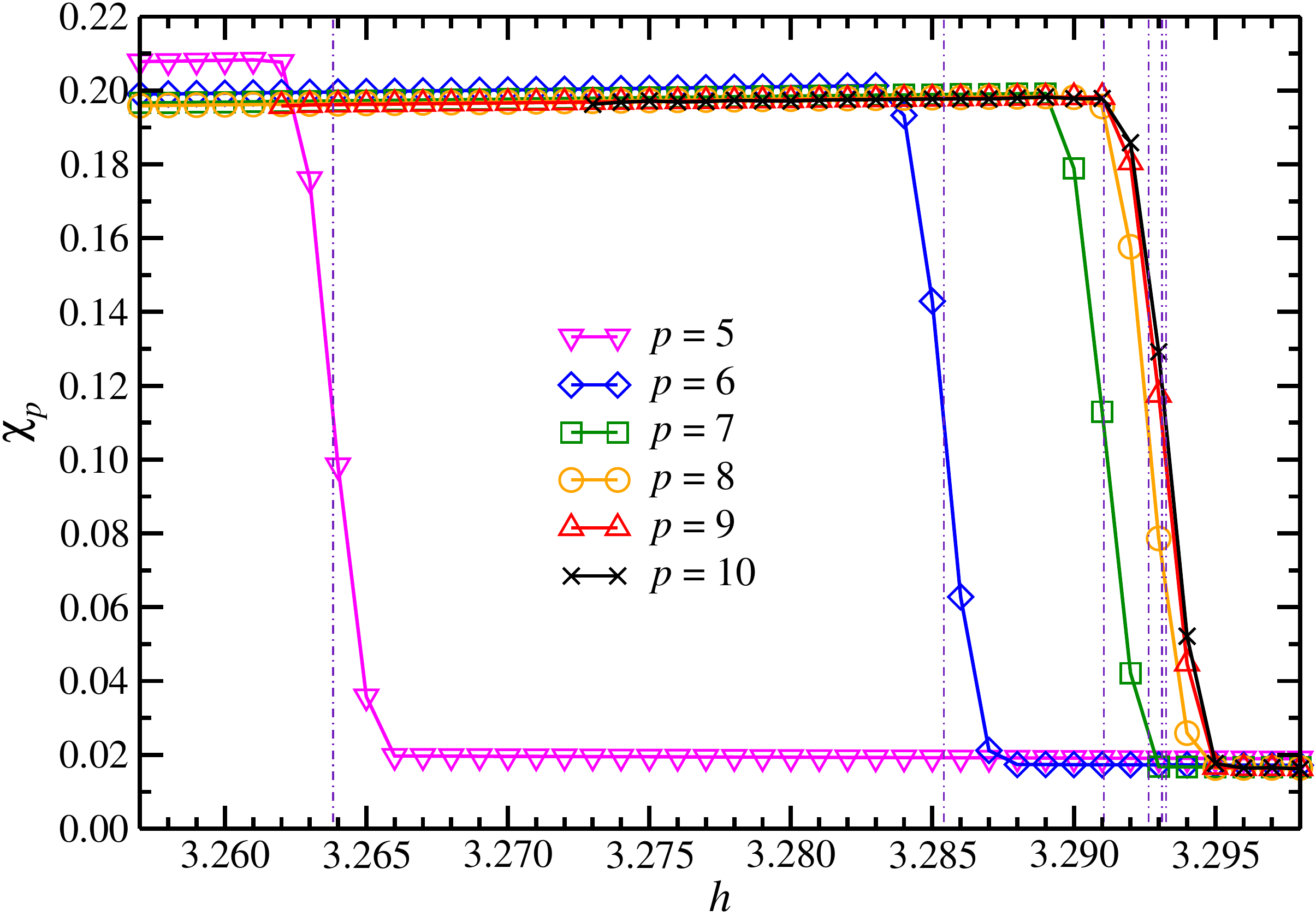}}
\caption{The magnetic susceptibility $\chi_p$ of the TFIM as a function of the magnetic field $h$ for the hyperbolic lattices with $p \in \{5, \dots,10\}$. The vertical dot-dashed lines serve as guides for the eye and correspond to the phase transitions $h_t^{(p)}$.}
\label{Fig6}
\end{figure}

Another independent way of obtaining (and confirming) the phase transition fields $h_t^{(p)}$ can be carried out by analyzing the magnetic susceptibility 
\begin{equation}
\chi_p=-\frac{\partial^2 E_0^{(p)}}{\partial h^2}\, .
\label{chi}
\end{equation}
The functional dependence of the susceptibility on the magnetic field $h$ is shown in figure~\ref{Fig6}. A non-diverging discontinuity of $\chi_p$ occurs at the identical phase transition fields $h_t^{(p)}$, which we have determined above by the spontaneous magnetization analysis and are depicted by the vertical dot-dashed lines. The inaccuracy comes from performing the second derivative in~\eref{chi} numerically, and the additional improvement rests in decreasing the spacing interval $\delta h$, i.e, in shrinking the distance between the magnetic fields, at which the ground-state energy is evaluated by TPVF. In the limit $\delta h \to 0$, the magnetic susceptibility undergoes a discontinuous jump at $h_t^{(p)}$~\cite{DaniskaGendiar}. It is obvious that there is no significant difference between the phase transition magnetic fields $h_t^{(p)}$ obtained by the analysis of the transverse magnetization $\langle S_p^{z}\rangle$ and the magnetic susceptibility $\chi_p$.     

Except for the analysis of the phase transition by the spontaneous magnetization $\langle S_p^{z}\rangle$ and the magnetic susceptibility $\chi_p$, the field dependence of the set of the optimal free variational parameters $W^*_p(\theta_j)$ also provides helpful information about the phase transition $h_t^{(p)}$. The pairs of the optimal variational parameters $W^*_p(\theta_j)$ coupled by spin-inversion symmetry smoothly collapse onto a single curve exactly at the phase transition for all considered lattice geometries. This process follows the identical behavior as we have presented in~\cite{DaniskaGendiar}. However, due to the large number of the variational parameters $N_{\rm Ising}^{(p)}$, we do not plot the $h$-dependence of $W^*_p(\theta_j)$ since the behavior remains qualitatively unchanged.
          
\section{Conclusions}

We have investigated three quantum spin-$\frac{1}{2}$ models (Heisenberg, XY, and TFIM) on a series of hyperbolic lattices by means of the numerical algorithm Tensor Product Variational Formulation~\cite{DaniskaGendiar}. The series of lattices is constructed by tessellation of regular $p$-sided polygons with the fixed coordination number equal to four, where $p \in \{5, \dots,11\}$. The Euclidean square lattice ($p=4$) has been also considered as a reference lattice, although we have discussed in~\cite{DaniskaGendiar} that TPVF applied to the models on the square lattice is less accurate than on the hyperbolic lattices ($p>4$).

The ground-state energies $E_0^{(p)}$ of the XY and the Heisenberg models have been studied in the absence of magnetic field on the series of the regular lattices with $4\leq p \leq 11$. Since no spontaneous symmetry breaking occurs in the two models at $h=0$, the spin-inversion symmetry helps to accelerate the TPVF algorithm due to significant reduction of the number of the free variational parameters. The resulting dependence of the ground-state energy per bond $E_0^{(p)}$ on the lattice geometry $p$ differs considerably for the two models. While the energies $E_0^{(p)}$ of the XY model form a monotonically increasing and exponentially saturated sequence with increasing $p$, the Heisenberg model induces a saw-like dependence containing the separated upper (odd $p$) and the lower (even $p$) branches, both of them converging exponentially fast to the common asymptotic value $E_0^{(\infty)}$ which corresponds to the ground-state energy on the Bethe lattice with the coordination number four.

We have analyzed the phase transition magnetic fields $h_t^{(p)}$ of the TFIM by the expectation value of the spontaneous magnetization $\langle S_p^z \rangle$, the associated magnetic exponent $\beta_p$, and the magnetic susceptibility $\chi_p$. We have calculated a sequence of the phase transition magnetic fields $h_t^{(p)}$, which is a strictly monotonous and increasing function, which converges exponentially to the asymptotic value $h_t^{(\infty)}$. This feature is completely analogous to a fast exponential saturation of the critical temperatures $T_c^{(p)}$ we had observed for the classical Ising model on the identical series of hyperbolic lattices in our earlier studies~\cite{hctmrg-Ising-5-4,hctmrg-Ising-p-4}. However, we have not found physical interpretation of this phenomenon yet. The quantum spin systems (as well as the classical ones) investigated on the hyperbolic lattices belong to the mean-field universality class, since infinite Hausdorff dimension of the hyperbolic lattice geometry exceeds the critical lattice dimensions $D_c = 3$ (for quantum models) and $D_c=4$ (for the classical ones). The linearity of the squared magnetization in the vicinity of the phase transition confirms the mean-field-like behavior, in which the associated magnetic exponents $\beta_p = \frac{1}{2}$. 

Although the set of the calculated phase transition magnetic fields $h_t^{(p)}$ and the ground-state energies $E_0^{(p)}$ are restricted to $4 \leq p \leq 11$, which is far away from the asymptotics $p\to\infty$, the fast convergence and the exponential character of $h_t^{(p)}$ and $E_0^{(p)}$ with increasing $p$ enables to estimate the respective quantities of the quantum spin models on the Bethe lattice ($p \to \infty$). In particular, we conjecture that the phase transition field of the TFIM on the Bethe lattice is positioned at $h_t^{(\infty)} = 3.29332$ and the ground-state energies per bond of the XY and the Heisenberg models, respectively, occur at $E_0^{(\infty)} = -1.08083446 $ and $-1.291944$. The three quantum spin models have not yet been considered on the Bethe lattice with the coordination number four.

\ack
The support received from the Grants QIMABOS APVV-0808-12 and VEGA-2/0130/15 is acknowledged.

\end{document}